\documentclass[11pt,a4paper]{article} 
\usepackage{jcappub}

\usepackage{amssymb}
\usepackage{amsmath}
\usepackage{amssymb}
\usepackage{natbib}
\usepackage{times}
\def\beq{\begin{equation}}
\def\enq{\end{equation}}

\begin{document}




\title{A method for evaluating 
the expectation value of a power spectrum using the probability density function of phases}


\author{G. A. Caliandro$^{1}$, D. F. Torres$^{1,2}$, \& N. Rea$^1$}

\affiliation{
$^{1}$Institute of Space Sciences (IEEC-CSIC), Campus UAB, Fac. de Ci\`encies, Torre C5, parell, 2a planta
08193 Barcelona,  Spain\\
$^{2}$Instituci\'o Catalana de Recerca i Estudis Avan\c{c}ats (ICREA) Barcelona, Spain}

\abstract{
%
Here, we present a new method to evaluate the expectation value of the power spectrum of a time series. A statistical approach is adopted to define the method. After its demonstration,
it is validated showing that it leads to the known properties of the power spectrum when
the time series contains a periodic signal. The approach is also validated in general with numerical simulations. 
The method puts into evidence the importance that is played by the probability density function of the phases associated to each time stamp for a given frequency, and how this distribution can be perturbed by the uncertainties of the parameters in the pulsar ephemeris. 
%
%
We applied this method to solve the power spectrum in the case the first derivative of the pulsar frequency is unknown and not negligible. We also undertook the study of the most general case of a blind search, in which both the frequency and its first derivative are uncertain. We found the analytical solutions of the above cases invoking the sum of Fresnel's integrals squared.
}

\keywords{stars: neutron, pulsars, gamma-rays: observations, time series, power spectrum}

\maketitle

\section{Introduction}

Pulsar timing from long observations is a current topic of astrophysics. 
This is especially due to the discovery of an unprecedented population of gamma-ray pulsars by the Large Area Telescope (LAT) on board the {\it Fermi} satellite \citep{psrCatalog}. 
The low rate of gamma-rays implies the need of very long observations so as to reach a significant detection of the pulsations. The survey mode of {\it Fermi}-LAT perfectly matches this need, and a timing consortium of radio and X-ray observers has been formed in order to build long ephemeris  to adequately extract pulsations \citep{TimeCons}.
More recently, adapting the radio techniques, \cite{PRay} presented a method to build long ephemeris using gamma-ray data. Since {\it Fermi}-LAT continuously monitors all the sky,  these ephemeris have an accuracy that is even better than the radio ones for  bright gamma-ray pulsars.
Efforts were also spent in order to find new techniques to  make the blind search for gamma-ray pulsations more efficient and computationally feasible. Indeed, due to the long integration times, a full Fourier analysis is computationally too demanding
for the search of pulsations in gamma-rays, particularly when lacking knowledge from other wavelengths (a blind search). 
A sub-population of radio quiet gamma ray pulsars (\cite{CTA1}, \cite{blindpsr}) has been discovered with LAT data thanks to the `time differencing technique'  \cite{Atwood2006}.
Recently, a new technique dedicated especially to the search of millisecond pulsars has also been developed \cite{Pletsh}.

The cited works on pulsar timing have greatly improved the analysis of isolated pulsars, and their blind search on gamma-rays. On the other hand, the search of gamma-ray pulsars in binary systems is performed only when their radio ephemeris are known. Indeed, the first pulsar catalog of {\it Fermi}-LAT \citep{psrCatalog} includes only five radio loud millisecond pulsars in binary systems. The reason is mainly that the Doppler shift due to the orbital motion of the pulsar wash out the pulsations at the observer's frame of reference. The correction for the orbital motion (as well as for other effects due to the companion star) need a very precise knowledge of the orbit. Since there are several parameters that define the orbit, uncertainties in one or more of them can vanish the effect of the correction. 
We have recently presented a systematic study on how the uncertainties of the orbital parameters affect the pulsed signal \cite{Cali12}. In that paper we already made use of a new method to compute the expectation value of the power spectrum at the signal and nearby frequencies, based on a statistical approach, which we will detail and prove here. 
The rest of the paper is organized as follows: In Section 2 we recall the main approaches used in the literature to compute the expectation value of a time series up to now.
In Section 3 we demonstrate our method, which we validate it in Section 4 by means of numerical  simulations and by reproducing known features of the power spectrum of a periodic times series.  
In sections 5 and 6, respectively, we apply the method to analytically solve the power spectrum in the case the first derivative of the frequency is uncertain and not negligible, and in the case of a blind search where both the frequency and its derivative are unknown.
Finally, in Section 7 we conclude underlining some interesting aspects of this approach.

\section{Context}

In our work we calculate the expectation value of the power spectrum following a statistical approach. 
Specifically, we treat the phase associated to each time stamp as a random variable, and the power spectrum as a function of it.
Therefore, starting from the probability density function of the phases, we make use of the known rules of statistics in order to derive the properties of the power spectrum.
We will show that with this approach we reach the same results obtained by other authors in different ways. But the novelty of our approach is that we introduce a different concept of the event phases and a different way  to treat them. We found that the approach we introduce can be particularly useful to solve specific problems that can arise in a timing analysis.
In order to highlight the difference of our method with respect to what can be found in literature, we first summarize other approaches in what follows.

\subsection{Lomb's approach}
In 1975, Lomb \cite{Lomb} undertook the study of the frequency analysis by means of the least-square method, which consists in fitting  sine waves 
to the data ($y_i$, taken at times $t_i, i=1,2,...,n$) in the time domain, 
and plotting versus the frequency ($\nu$) the reduction ($\Delta R$) in the sum of the residual squared respect to fitting a constant value (see also \cite{Barning})
\beq
\Delta R = \sum_i (y_i - c)^2 - \sum_i \left[ y_i - (a\,{\rm cos} (2\pi \nu t_i) + b\,{\rm sin} (2\pi \nu t_i) +c)\right] ^2.
\label{ls}
\enq
Lomb \cite{Lomb} (see his Section 2) found the formula of the least-squared spectrum (a posteriori called Lomb-Scargle periodogram) and demonstrates that the formula of the classical periodogram is an approximation for it. From this base the statistical properties of the power spectrum of random noise were examined, as well as the effect of noise on the spectrum of a sine wave.

\subsection{van der Klis' approach}

van der Klis \cite{Klis} (see his Section 2.5) explained the relation between the continuous Fourier transform and the Discrete Fourier Transform (DFT) introducing the concepts of windowing and sampling. 
The Fourier transform of an infinitely extended continuous function $x(t)$ ($-\infty <t< \infty$) is defined as
\beq
a(\nu) = \int_{-\infty}^{\infty} x(t) e^{2\pi i \nu t} dt.
\label{Fourier}
\enq
When a real instrument is used to detect the phenomena described by the function $x(t)$, we need to take into account two effects. The observation will be finite in time, and the instrument hardware will detect discrete samples of the function $x(t)$. In other words, what we detect is the function $x(t)$ multiplied by both the window function, which describe the structure and the duration of the observations, and by the characteristic sampling function of the instrument.
The simplest window function is
\begin{eqnarray}
\omega(t)=\begin{cases} 1, & \mbox{if }\, 0\leq t<T \\ 0, & \mbox{otherwise }
\end{cases},
\end{eqnarray}
where $T$ is the duration of the observation.
The sampling function is 
\beq
s(t) = \sum_{k=-\infty}^{\infty} \delta\left( t-\frac{kt}{N}\right),
\enq
where $N$ is the number of bins in the observation time $T$.
The observed function is then
\beq
\tilde{x}(t) = x(t) \cdot \omega(t) \cdot s(t).
\label{xtilde}
\enq
Substituting (\ref{xtilde}) into (\ref{Fourier}) we end up with the definition of the DFT.
A powerful theorem of the Fourier analysis states that the Fourier transform of the product of two functions is the convolution of the Fourier transforms of these two functions. So, if $a(\nu)$ is the Fourier transform of $x(t)$, and $b(\nu)$ that of $y(t)$, then the Fourier transform of $x(t)y(t)$ is $a(\nu)*b(\nu) \equiv \int_{-\infty}^{\infty} a(\nu')b(\nu - \nu')d\nu'$.
Making use of this theorem applied to Eq. \ref{xtilde}, van der Klis \cite{Klis} derived the main properties of the power spectrum calculated for an event data time series.

\subsection{Ransom et al.'s approach}

In an alternative way, Ransom et al. \cite{Ransom2002} (see their Section 2.1) derived the properties of the Fourier power of a time series from a direct study of the Discrete Fourier Transform. The $k^{th}$ element of a DFT of a uniformly-spaced time series $n_j$ ($j=0,1,....,N-1$) is defined as
\beq
A_k = \sum_{j=0}^{N-1} n_j e^{-2\pi ijk/N},
\enq
where for a time spacing $dt$, and a total observation length $T$, the frequency of the $k^{th}$ element is $f_k = k/T$, and the index $j$ indicate the time stamp $t=j dt$.
If the DFT summation is drawn on a complex plane, it appears as a simple vector addition with each element rotated by $-2\pi k/N$ from the previous element.
Starting from this representation of the DFT summation, \cite{Ransom2002} derived the Fourier response and its power to full noise time series as well as to periodic signals. 
\\

To describe the statistical properties of the power spectrum of a time series composed by noise and a periodic signal, both \cite{Klis}, and \cite{Ransom2002} follow the description given by Groth \cite{Groth}. He considers the Fourier power of a single frequency bin as a random variable. Invoking the central limit theorem, Groth \cite{Groth} demonstrates in a few steps that the Fourier power of a noisy time series has a $\chi^2$ distribution with 2 degrees of freedom. As one can notice, none of the authors cited above directly define the phase of the time stamps ($\theta = 2 \pi \nu t$), nor use it as a random variable, 
in contrast with what follows.

\section{Power spectrum expectation value}

\subsection{First approach}

Following the definition given in \cite{Scargle1982}, the power spectrum of a 
sample data set $\{X(t_i), i=1,2,....,N_0\}$ calculated at an arbitrary test frequency $\omega$ is given by
\begin{equation}
\! \! 
P(\omega) \!=\! \frac{1}{N_0} \!
\left[ \left( \sum_{i=1}^{N_0} X_i {\rm cos}(\omega t_i)\right) ^2
\! \!  + \!
\left( \sum_{i=1}^{N_0} X_i {\rm sin}(\omega t_i)\right) ^2\right] . \!
\label{2.1}
\end{equation} 
Here, we consider the series of the arrival times of single events on a detector, i.e. the sample data set $X$ is such that $X_i=1$ for each $i$.
Attributing a phase value to each event $\theta_i = \omega t_i$ we have for the power spectrum
\begin{equation}
P(\omega) = \frac{1}{N_0}\left[ \left( \sum_{i=1}^{N_0} {\rm cos}(\theta_i)\right) ^2+\left( \sum_{i=1}^{N_0} {\rm sin}(\theta_i)\right) ^2\right].
\label{pow}
\end{equation} 
This definition corresponds to the Rayleigh power, and it differs by just a factor 2 with respect to 
the $Z_1^2$ test introduced by \cite{Buccheri1983}.
The variable $Z_1^2$ has a probability density function 
(pdf) for the null hypothesis equal to that of a $\chi^2$ with 2 degrees of freedom, where null hypothesis is used to signify that there is no periodic signal at frequency $\omega$. 
In what follows we focus on how to calculate the expectation 
value of the power spectrum in Eq. (\ref{pow}) using the statistical 
properties of the phase distribution $P_{\theta}(\theta)$.

Here and throughout the paper $\theta$ does not refer to phase of the
pulsar, but to a phase relative to an arbitrary test frequency, as defined above.
$P_{\theta}(\theta)$ refers to the theoretical continuous distribution of the phases. 

For a large number of events ($N_0 \gtrsim 100$) in Eq. (\ref{pow}), 
the sums of the trigonometric functions of the phases, 
cos$(\theta_i)$ and sin$(\theta_i)$, are well approximated by
the expectation value of their continuous distributions times $N_0$. Thus,
%
\begin{equation}
\sum_{i=1}^{N_0} {\rm cos}(\theta_i) \longrightarrow N_0  \left\langle {\rm cos}(\theta_i) \right\rangle 
:= N_0\int_{-1}^1 {\rm cos}(\theta) \cdot P_{\rm cos}(\theta) d{\rm cos}(\theta),
\label{2.3}
\end{equation}
\begin{equation}
\sum_{i=1}^{N_0} {\rm sin} (\theta_i) \longrightarrow N_0  \left\langle {\rm sin}(\theta_i) \right\rangle  
:= N_0\int_{-1}^1 {\rm sin} (\theta) \cdot P_{\rm sin} (\theta) d{\rm sin}(\theta)  ,
\label{2.4}
\end{equation}
where $P_{\rm cos}(\theta)$ and $P_{\rm sin}(\theta)$ are the continuous distributions of ${\rm cos}(\theta)$ and ${\rm sin}(\theta)$ expressed as function of $\theta$, respectively. 
Substituting these in Eq. (\ref{pow}), the power spectrum is given by
\begin{equation}
P(\omega) = N_0\left[ \left\langle {\rm cos}(\theta_i) \right\rangle^2 + \left\langle {\rm sin}(\theta_i) \right\rangle^2\right] .
\label{pow2}
\end{equation} 
Since we used the mean values of cosine and sine, Eq. (\ref{pow2}) is the expectation value of the power spectrum at the frequency $\omega$.

\subsection{A more rigorous approach}

This can be demonstrated more rigorously in the following way. Take a set of $N$ independent random variables $z_i$, all with the same variance $\sigma$ and the same average $\left\langle z \right\rangle$. The expectation value of their squared sum is
\begin{eqnarray}
&& \left\langle \left( \sum z_i \right)^2    \right\rangle   
=\left\langle \sum z_i^2 + 2\sum_{i\neq j}z_i z_j \right\rangle =
\nonumber \\ &&
\sum \left\langle z_i^2 \right\rangle + 2\sum_{i\neq j} \left\langle z_i \right\rangle \left\langle z_j \right\rangle= 
\nonumber \\ &&
 \sum \left( \left\langle z_i^2 \right\rangle - \left\langle z_i \right\rangle^2 \right) + 2\sum_{i\neq j} \left\langle z_i \right\rangle \left\langle z_j \right\rangle + \sum  \left\langle z_i \right\rangle^2= 
\nonumber \\&&
 N\sigma^2 + 2\left( N^2 - \sum_{i=1}^N i \right) \left\langle z \right\rangle^2 + N \left\langle z \right\rangle^2 = 
\nonumber \\ &&
 N\sigma^2 + N^2 \left\langle z \right\rangle^2 \xrightarrow{N\rightarrow\infty, \left\langle z \right\rangle\neq 0} N^2 \left\langle z \right\rangle ^2
\label{proof}
\end{eqnarray}
where we used the commutative properties of the average operator ($\left\langle \right\rangle $), and we expressed the number of the double products from the square of a sum as $\sum_{i\neq j} 1 = \left( N^2 - \sum_{i=1}^N i \right)$.
Averaging Eq. (\ref{pow}), and doing  algebra as above on $\left( \sum {\rm cos}(\theta_i)\right) ^2$ and $\left( \sum {\rm sin}(\theta_i)\right) ^2$, it can be proven that the expectation value of the power spectrum is given by Eq. (\ref{pow2}). 

In the last step of Eq. (\ref{proof}) it is shown that $\left\langle \left( \sum z_i \right)^2  \right\rangle = N^2 \left\langle z \right\rangle^2$ only if $\left\langle z \right\rangle\neq 0$, whereas in the opposite case, the term with the variance ($N\sigma^2$) is dominant.
This represents a limitation of our method, which is correct only for frequencies $\omega$ which are close to the signal ($\omega_0$). Indeed, for frequencies far from the signal, the distribution of the phases is flat and the power spectrum is dominated by noise.
Several works dedicated to the study of the noise in the power spectrum can be found in literature (e.g. \cite{Scargle1981}). Here, we focus on the power spectrum close to the signal peak, with the aim of using 
the results for a future development of optimized software dedicated to the search of periodic signals.

\subsection{Distributions}

In the next paragraphs we will find a formula for $P_{\rm cos} (\theta)$ and $P_{\rm sin} (\theta)$, in order to solve the integrals in Eqs. (\ref{2.3}) and (\ref{2.4}), and calculate Eq. (\ref{pow2}). 
We shall use that the probability density function (pdf) of a variable $z=f(x)$ that is function of a random variable $x$ whose pdf $P_x(x)$ is known, can be calculated as 
\begin{equation}
P_{z}(z) = \sum_i \frac{P_x(x_{2\_i})}{f^{\prime}(x_{2\_i})} - \frac{P_x(x_{1\_i})}{f^{\prime}(x_{1\_i})}
\label{rotondi}
\end{equation}
(Eq. \ref{rotondi} is taken from \cite{Rotondi}. It can also be found in e.g. \cite{Miller}). 
Here, the prime (as in $f^{\prime}$) represents a derivative with respect to $x$, and the intervals $\left[ x_{1\_i}, x_{2\_i}\right] $ are those for which for a given $z=z_0$, $f(x)<z_0$. At the first or the last interval (with respect to the validity range of $x$),  it could happen that the edges, $x_{1\_i}$ and $x_{2\_i}$, respectively, can not be properly defined. 
The corresponding term in Eq. (\ref{rotondi}) is null in these cases. 

In our case, $P_z$ in Eq. (\ref{rotondi}) corresponds to $P_{\rm cos}$ or $P_{\rm sin}$, while $P_x$ is the phase distribution $P_{\theta}$.
The phases $\theta=\omega t$ are defined in the full range $(0, \omega T_{\rm obs})$, where $T_{\rm obs}$ is the total length of the observation. We will find useful to express $\omega T_{\rm obs}$ as
\begin{equation}
\omega T_{\rm obs} = 2\pi N  + R 
\label{11}
\end{equation}
where $N$ is the number of entire cycles of $2\pi/\omega$ contained in the observation time $T_{\rm obs}$ and $R$ is the fractional part of the last cycle ($R = {\rm fmod}(\omega T_{\rm obs}, 2\pi)$) expressed in phase.

\subsection{$P_{\rm cos}$}

\begin{figure}[t]
\center
\includegraphics[width=0.8\columnwidth]{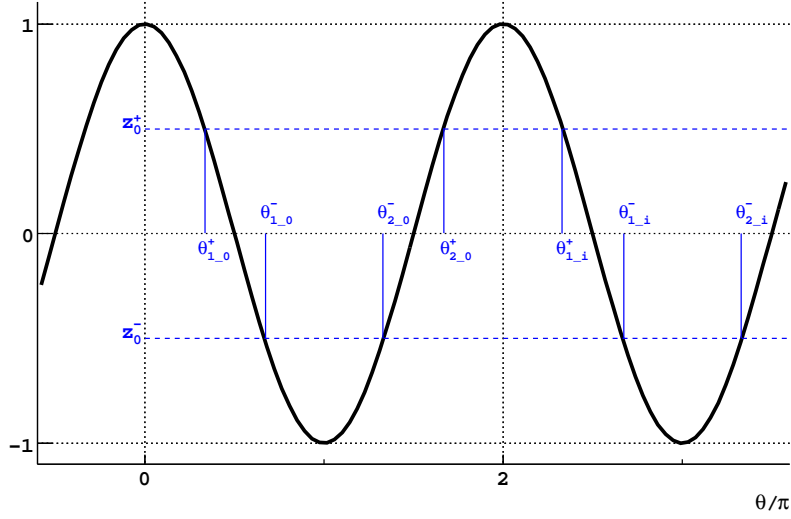}
\caption{The function $z={\rm cos}(\theta)$. 
The edges of the intervals for which cos$(\theta) < z_0^+$ are marked with $\theta^{+}_{1\_i}$, and $\theta^{+}_{2\_i}$, where the subindex $_i$ indicates the cycle number of the sine function.
Similarly, the edges of the intervals for which cos$(\theta) < z_0^-$ are marked with $\theta^{-}_{1\_i}$, and $\theta^{-}_{2\_i}$.\label{cos}}
\end{figure}

We shall start with the calculation of $P_{\rm cos}$. Figure \ref{cos} shows the cosine function. We marked a generic value $z_0^+>0$ and  found the edges of the intervals for which cos$(\theta) < z_0^+$. They are marked in the figure as $\theta^{+}_{1\_i}$, and $\theta^{+}_{2\_i}$, where the subindex $_i$ indicates the cycle number of the cosine function. In the same way we marked a generic value lower than 0 ($z_0^-<0$), and the respective intervals $\left[ \theta^{-}_{1\_i}, \theta^{-}_{2\_i} \right] $ for which cos$(\theta) < z_0^-$. 
 Since the codomain of the inverse function ${\rm acos}(z)$ is $[0, \pi]$, it is convenient to express the edges of the intervals marked in figure \ref{cos} in the following way
\begin{align}
  &\theta^{+/-}_{1\_0} = {\rm acos}(z_0), \nonumber \\
  &\theta^{+/-}_{2\_0} = 2\pi - \theta^{+/-}_{1\_0} , \\
  &\theta^{+/-}_{1\_i} = 2\pi i + \theta^{+/-}_{1\_0}, \nonumber \\
  &\theta^{+/-}_{2\_i} = 2\pi (i+1) - \theta^{+/-}_{1\_0} .\nonumber 
\end{align}
Substituting these values in Eq. (\ref{rotondi}), the most general expression for the pdf of ${\rm cos}(\theta)$ is
\begin{eqnarray}
P_{\rm cos}(\theta) = \frac{1}{{\rm sin}(\theta)}\left\lbrace \sum_{i=0}^{N-1}\left[ P_{\theta}(2\pi i + \theta) + \right. \right.
 \left. \left. 
 P_{\theta}(2\pi(i+1) - \theta) \right]_{0,\pi}  \right\rbrace + B(\theta; R).
 \label{Pcos}
\end{eqnarray}
Here we have substituted $\theta^{+/-}_{1\_0}$ with $\theta$, which for this expression is constrained in the range $[0, \pi]$, as is specified in the formula at the feet of the closing of the square bracket. In Eq. (\ref{Pcos}), sin$(\theta)$ is the derivative of the cosine calculated at $(2\pi i + \theta)$ and $(2\pi (i+1) - \theta)$.
The sum is over the number $N$ of entire cycles contained within the observation (see Eq. \ref{11}), while the term $B(\theta; R)$ takes into account the fractional part of the last cycle. 
This term is in general negligible, because commonly one would have $N \gg 1$. 
It strongly depends on how long is the fraction of the last cycle, thus, on $R$, defined by Eq. (\ref{11}).  
For example, if $R<\pi$ the interval for which cos$(\theta) < z_0$ is $\left[ \theta_{1\_N}, R \right] $. Since the second edge of the interval is not properly defined, but it is set to the end of the observation, the term $P_x(x_{2\_N})/f^{\prime}(x_{2\_N})$ in Eq. (\ref{rotondi}) has to be set to zero. Finally, $B(\theta; R)$ is equal in this case to $[P_{\theta}(2\pi N + \theta)/{\rm sin}(\theta)]_{0,R}$, where the subindices indicate the range of validity of $\theta$ only in this last cycle.

\subsection{$P_{\rm sin}$}

\begin{figure}
\center
\includegraphics[width=0.8\columnwidth]{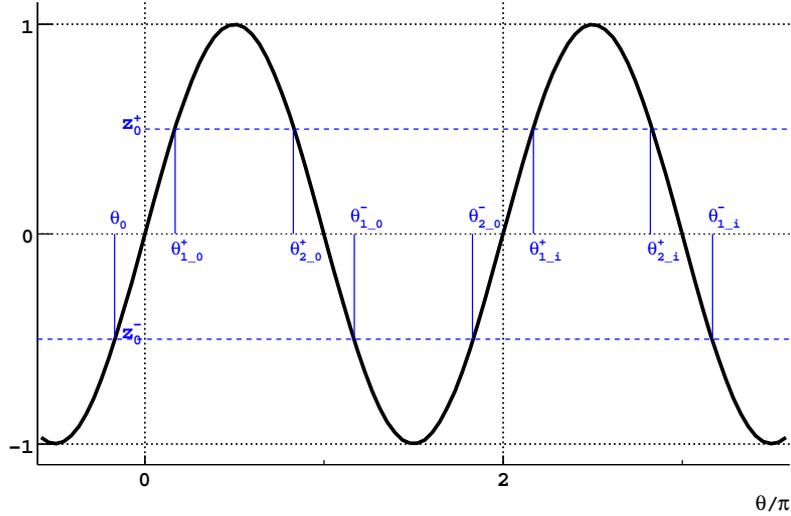}
\caption{The function $z={\rm sin}(\theta)$. 
The edges of the intervals for which sin$(\theta) < z_0^+$ are marked with $\theta^{+}_{1\_i}$, and $\theta^{+}_{2\_i}$, where the subindex $_i$ indicates the cycle number of the sine function.
Similarly, the edges of the intervals for which sin$(\theta) < z_0^-$ are marked with $\theta^{-}_{1\_i}$, and $\theta^{-}_{2\_i}$. The phase $\theta_{0}$ is useful in the computations, as described in the text.\label{sin}}
\end{figure}

The solution for $P_{\rm sin} (\theta)$ is similarly found.
Figure \ref{sin} plots the sine function. We marked there a generic value $z_0^+>0$, and the edges of the intervals for which sin$(\theta) < z_0^+$ ($\theta^{+}_{1\_i}$ and $\theta^{+}_{2\_i}$, where the subindex $_i$ indicates the cycle number of the sine function). In the same way we marked a generic value lower than 0 ($z_0^-<0$), and the respective intervals $\left[ \theta^{-}_{1\_i}, \theta^{-}_{2\_i} \right] $ for which sin$(\theta) < z_0^-$. Since in this work $\theta=\omega t$ is positively defined, the phase $\theta_0$ also marked in the figure is not a solution of interest for us, but we need it because the inverse function ${\rm asin}(z)$ has the codomain $[-\pi/2, \pi/2]$. 
The intervals defined for $z_0^+>0$ are symmetric respect to $\pi/2 + 2\pi i$, while the intervals defined for $z_0^-<0$ are symmetric respect to $\frac{3}{2}\pi + 2\pi i$. For this reason it is not immediate to join the two cases. Indeed, we will split them defining $P^+_{\rm sin}$ and $P^-_{\rm sin}$.
For positive values of $z_0$, the edges of the intervals can be expressed as
\begin{align}
  &\theta^{+}_{1\_0} = {\rm as in}(z_0^+)        ,     \nonumber\\
  &\theta^{+}_{2\_0} = \pi - \theta^{+}_{1\_0}   ,    \\
  &\theta^{+}_{1\_i} = 2\pi i + \theta^{+}_{1\_0} ,    \nonumber\\
  &\theta^{+}_{2\_i} = 2\pi(i+1/2) - \theta^{+}_{1\_0} .  \nonumber
\end{align}
Substituting in Eq. (\ref{rotondi}), we get
\begin{eqnarray}
P^+_{\rm sin}(\theta) =  \frac{1}{{\rm cos}(\theta)}\left\lbrace \sum_{i=0}^{N-1}\left[ P_{\theta}(2\pi i + \theta) + 
\right. \right. 
\left. \left.
 P_{\theta}(2\pi (i + 1/2) - \theta) \right]_{0,\frac{\pi}{2}}  \right\rbrace ,
\label{Psin+}
\end{eqnarray}
where here $\theta=\theta^{+}_{1\_0}$ is defined in the range $\left[ 0,\pi/2 \right] $, as indicated by the subindices of the closing square bracket.
For negative $z_0$, the edges of the intervals are
\begin{align}
& \theta_0 = {\rm as in}(z_0^-) ,  \nonumber\\
& \theta^{-}_{1\_0} = \pi - \theta_0  ,   \nonumber\\
& \theta^{-}_{2\_0} = 2\pi + \theta_0 ,  \\
& \theta^{-}_{1\_i} = \pi(i+1/2) - \theta_0  ,  \nonumber\\
& \theta^{-}_{2\_i} = 2\pi(i+1) + \theta_0  .  \nonumber
\end{align}
Substituting in Eq. (\ref{rotondi}), we obtain
\begin{eqnarray}
P^-_{\rm sin}(\theta) =  \frac{1}{{\rm cos}(\theta)}\left\lbrace \sum_{i=0}^{N-1}\left[ P_{\theta}(2\pi(i+1) + \theta) + \right. \right. 
\left. \left. P_{\theta}(2\pi (i + 1/2) - \theta) \right]_{-\frac{\pi}{2},0}  \right\rbrace ,
\label{Psin-}
\end{eqnarray}
where here $\theta=\theta_0$ is defined in the range $\left[-\pi/2, 0 \right] $, as indicated by the subindices of the closing square bracket.
In Eqs. (\ref{Psin+}) and  (\ref{Psin-}) cos$(\theta)$ at the denominator is the derivative of the sine calculated at 
$(2\pi i+\theta)$, $(2\pi(i+1/2)-\theta)$, and $(2\pi(i+1)+\theta)$.

Finally, we can sum the Eqs. (\ref{Psin+}) and (\ref{Psin-}) to get the complete solution for $P_{\rm sin}$, 
\begin{eqnarray}
P_{\rm sin}(\theta) = && \frac{1}{{\rm cos}(\theta)}\sum_{i=0}^{N-1} \left\lbrace 
\left[ P_{\theta}(2\pi(i+1/2) - \theta) \right]_{-\frac{\pi}{2},\frac{\pi}{2}} + \right.  \left.
\left[ P_{\theta}(2\pi i + \theta) \right]_{0,\frac{\pi}{2}}  + \right.
\nonumber \\ && \left.
\left[ P_{\theta}(2\pi(i+1) + \theta) \right]_{-\frac{\pi}{2},0}  \right\rbrace 
+ B(\theta; R)
\label{Psin}
\end{eqnarray}
Also here the term $B(\theta; R)$ takes into account the fraction of the last cycle of the sine function, and it is in general negligible.

We have now all the ingredients to solve the integrals in Eqs. (\ref{2.3}) and (\ref{2.4}), and 
calculate the expectation value of the power spectrum in Eq. (\ref{pow2}).
In both, Eqs. (\ref{Pcos}) and (\ref{Psin}), a key role is played by the sum of the terms $\sum_{i=0}^{N-1} P_{\theta}(2\pi(i+k) \pm \theta)$, where $k = 0, 1/2, 1$, as well as the sign of $\theta$ depend on the cases. This corresponds to nothing more than the distribution of the phases folded by $2 \pi$. This will be commented in the last Section. The solutions we found here are free from any assumption on $P_{\theta}$, and in this sense they are universal.

\section{ Numerical validation}

In this section we are going to validate in two different ways the method formulated to evaluate the expectation value of the power spectrum. First, the results found for the pdf of the sine and cosine (Eqs. \ref{Psin}, \ref{Pcos}) will be checked with simulations. Secondly, we shall show that using our approach we can derive all the already known features of the power spectrum.

\subsection{Simulations}

\begin{figure*}
 \centering
\begin{minipage}{1.0\textwidth}
\centering
 \includegraphics[width=0.4\textwidth]{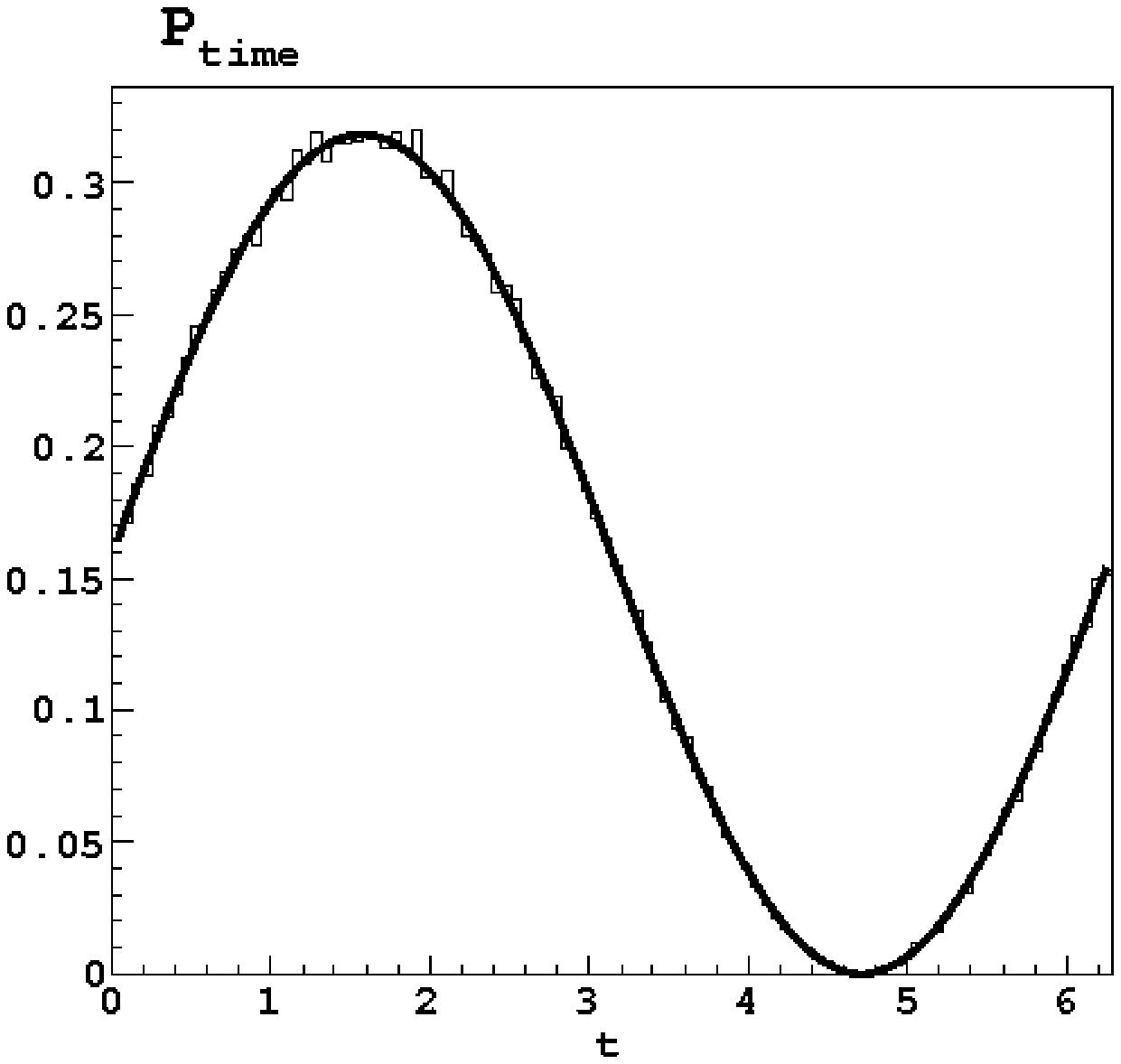}
 \hspace{0.5 cm}
 \includegraphics[width=0.4\textwidth]{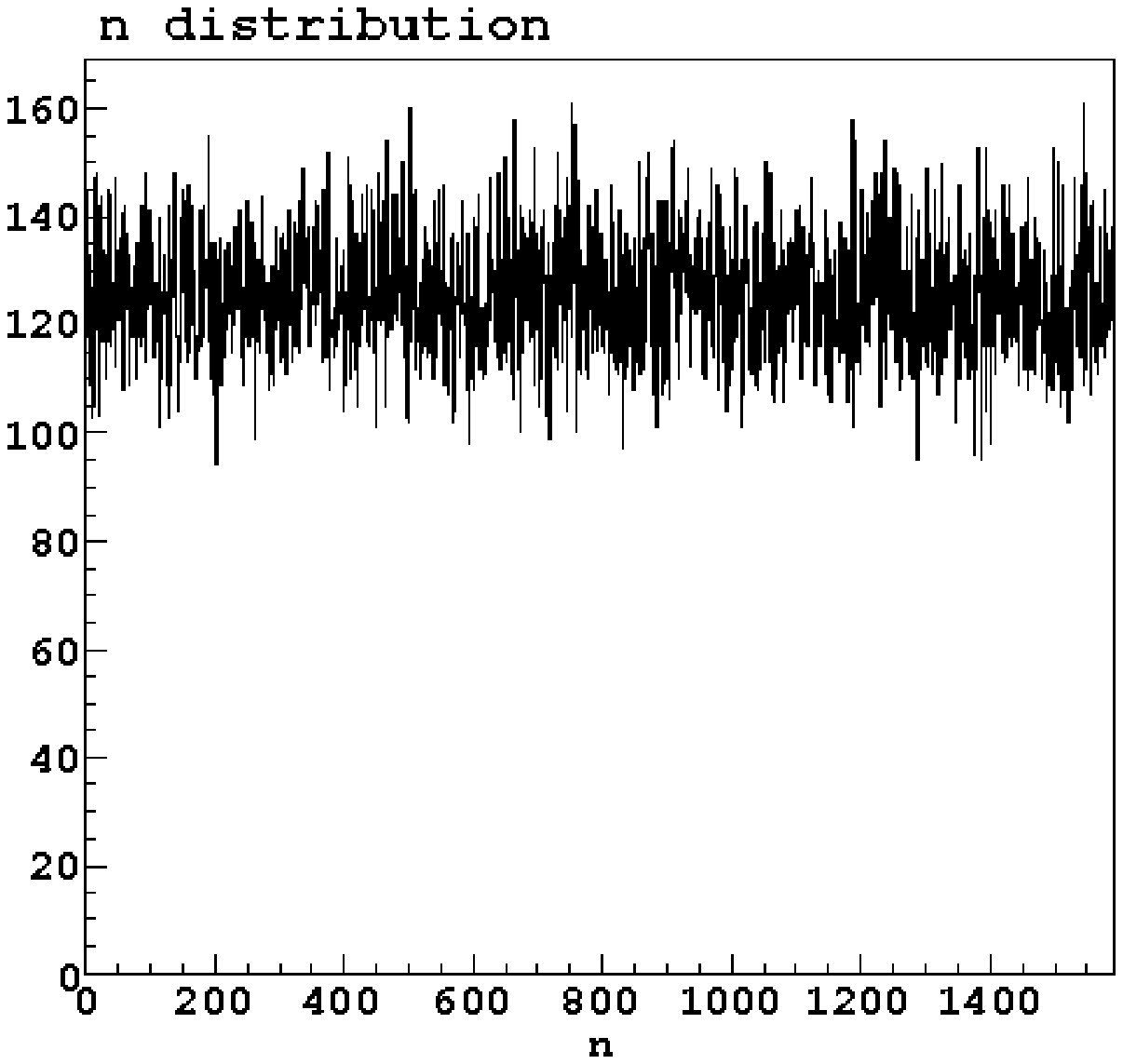}
\end{minipage}
\caption{Left panel: distribution of the time stamps after the first step in the simulation. Right panel: distribution of the integers n among $0$ and $\omega_0 T_{\rm obs}/2\pi$. 
\label{simulation}} 
\end{figure*}

In order to check the formulae we have found in the previous Section, we wrote a code to simulate the arrival time series from a sinusoidal signal of frequency $\omega_0$ (Eq. \ref{8}), rate $r$, observed for a time $T_{\rm obs}$ and having a distribution given by
\begin{equation}
 P_t(t) = 1+{\rm sin}(\omega_0 t).
 \label{8}
\end{equation}
The arrival time series is simulated in two steps. First, all the time stamps are simulated as random numbers in the range $0,2\pi/\omega_0$, following the distribution of Eq. (\ref{8}). Then, in order to cover the full duration of the observation, each time stamp has been randomly delayed adding a value $({2\pi}/{\omega_0}) n$, where $n$ is a random integer number uniformly distributed in $0, \omega_0 T_{\rm obs}/2\pi$.
Figure \ref{simulation} shows the distributions of the time stamps in the first step for $\omega_0=1.0$ s$^{-1}$ (left panel), and the distribution of $n$ for the second step, for $T_{\rm obs}=1.0$E+4 s (right panel). This procedure requires that the observation time $T_{\rm obs}$ is exactly an integer multiple of the signal period. This implies that for $\omega=\omega_0$ the term $R$ in equation \ref{11} is null, while it is in general not true for $\omega \neq \omega_0$.
Once the arrival time series has been fully simulated, we calculate the phases $\theta_i = \omega t_i$, their  cosine, and sine value, and finally the power at the frequency $\omega$ from Eq. (\ref{pow}).

\begin{figure*}
 \centering
\includegraphics[width=0.49\textwidth]{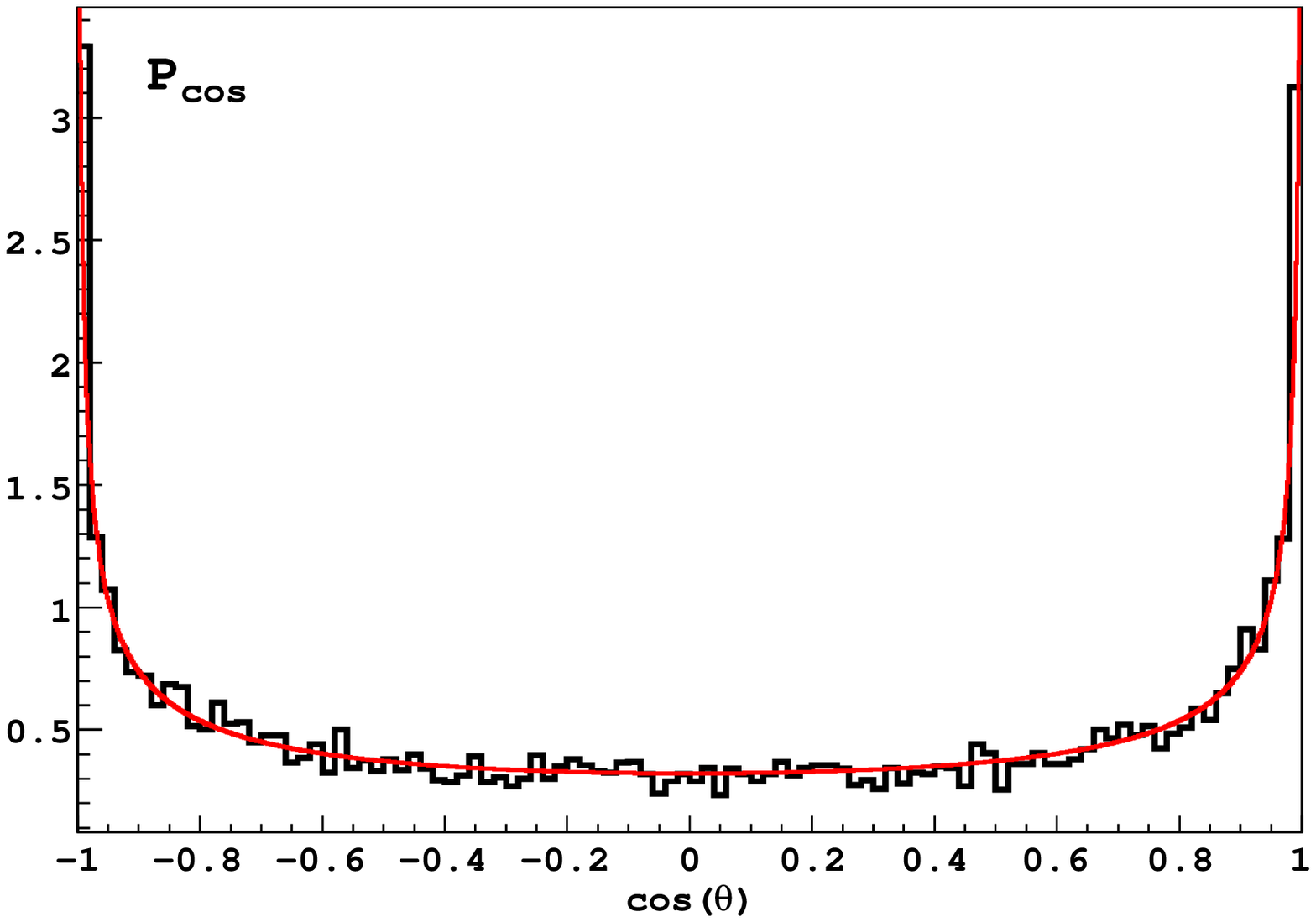}
\includegraphics[width=0.49\textwidth]{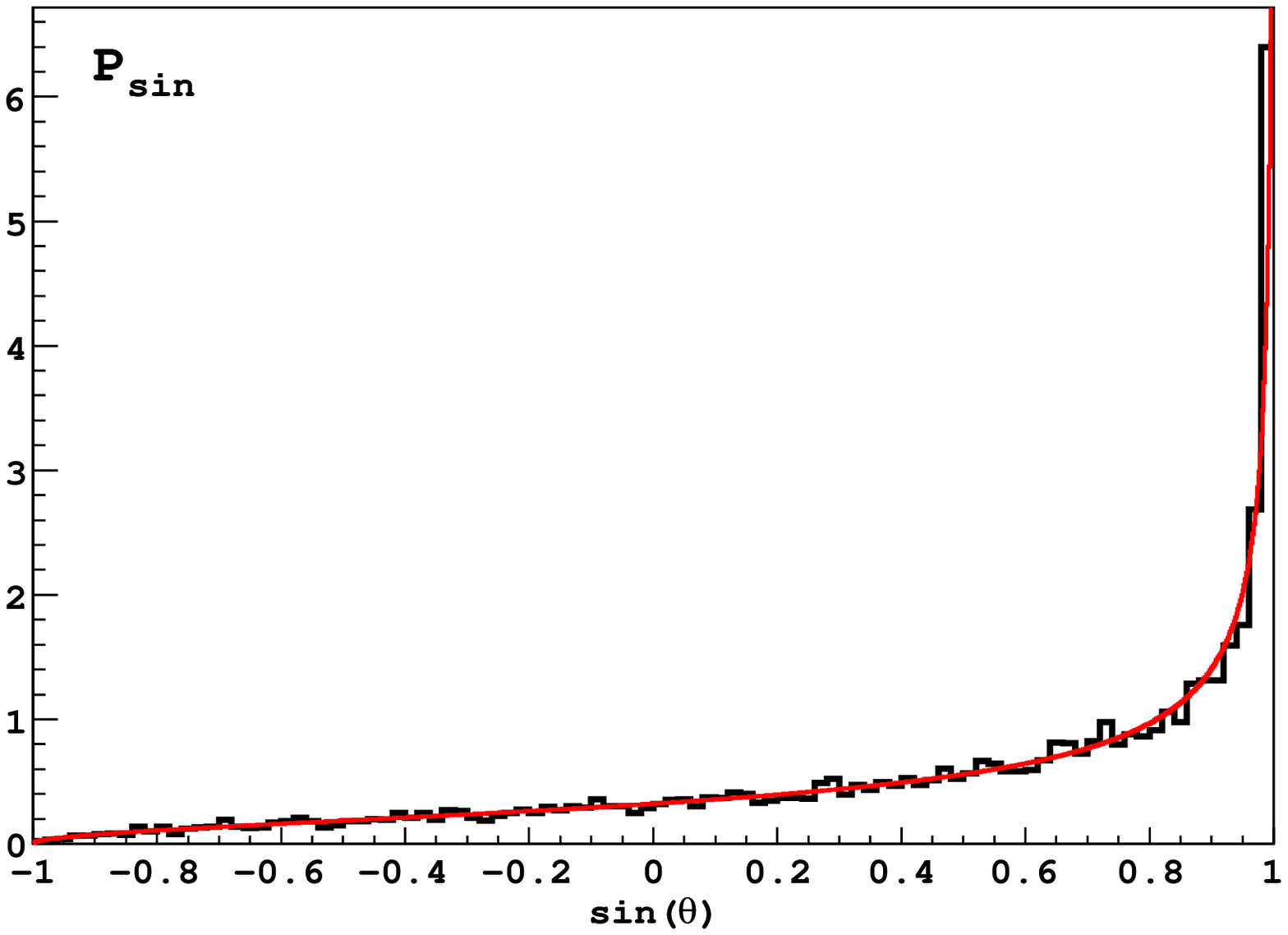}
 \caption{Distribution of sin$(\theta)$ (left), and cos$(\theta)$ (right) for the case $\omega=\omega_0$. 
The red lines represent the analytical formulae for the pdf. \label{sc@peak}}
 \end{figure*}

\begin{figure*}
 \centering
 \includegraphics[width=0.95\textwidth]{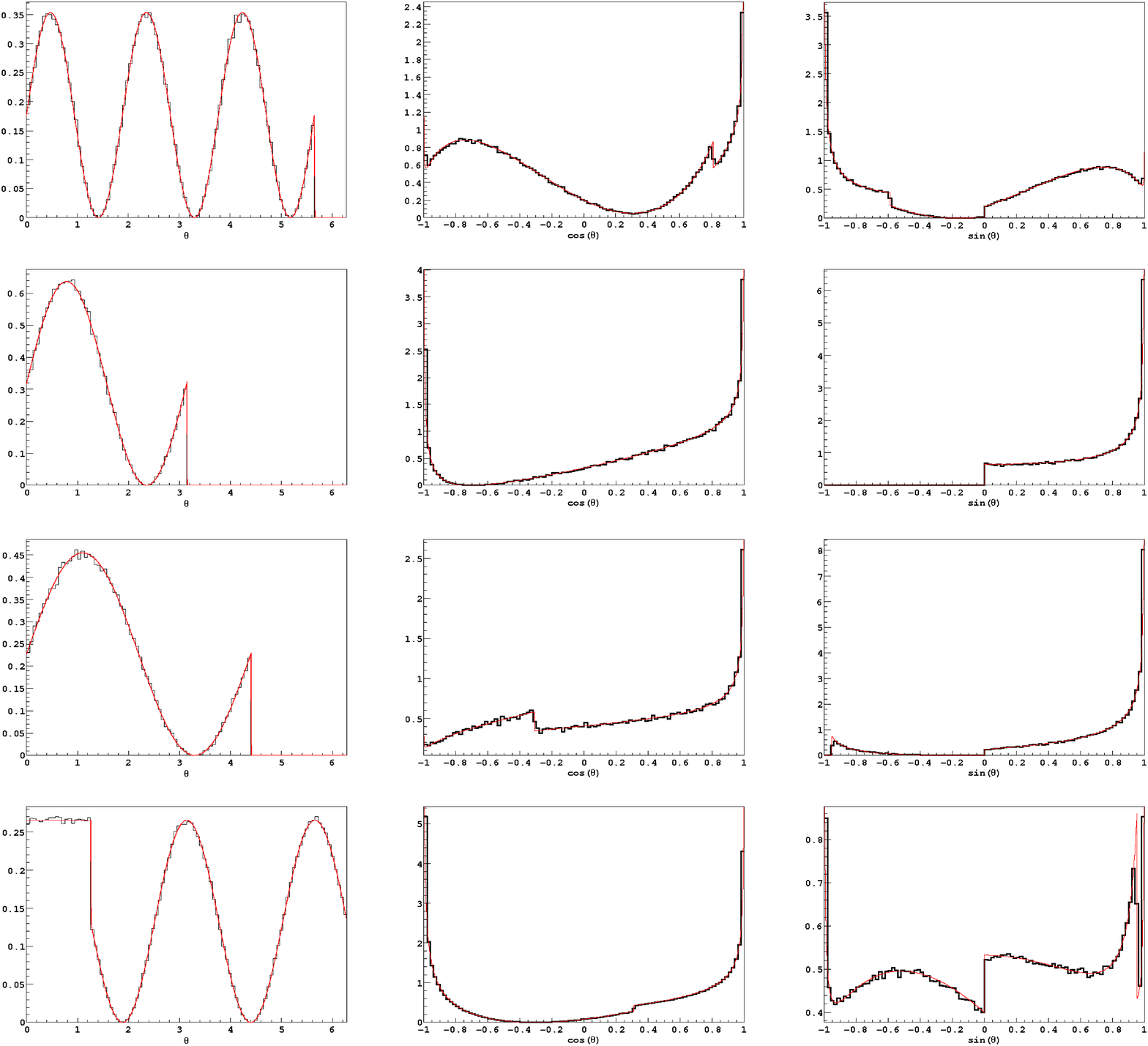}
 \caption{Distributions of $\theta$ (left), cos$(\theta)$ (middle), and sin$(\theta)$ (right) from the simulation. The analytical pdf is over imposed (red lines). First row: $\omega=0.3\omega_0$, $N=0$, $R=5.6$. 2nd row: $\omega=0.5\omega_0$, $N=0$, $R=3.14$. 3rd row: $\omega=0.7\omega_0$, $N=0$, $R=4.39$. 4th row: $\omega=0.4\omega_0$, $N=1$, $R=1.25$. \label{SimulationPanels}}
\end{figure*}

\begin{figure*}
 \centering
 \includegraphics[width=0.32\textwidth]{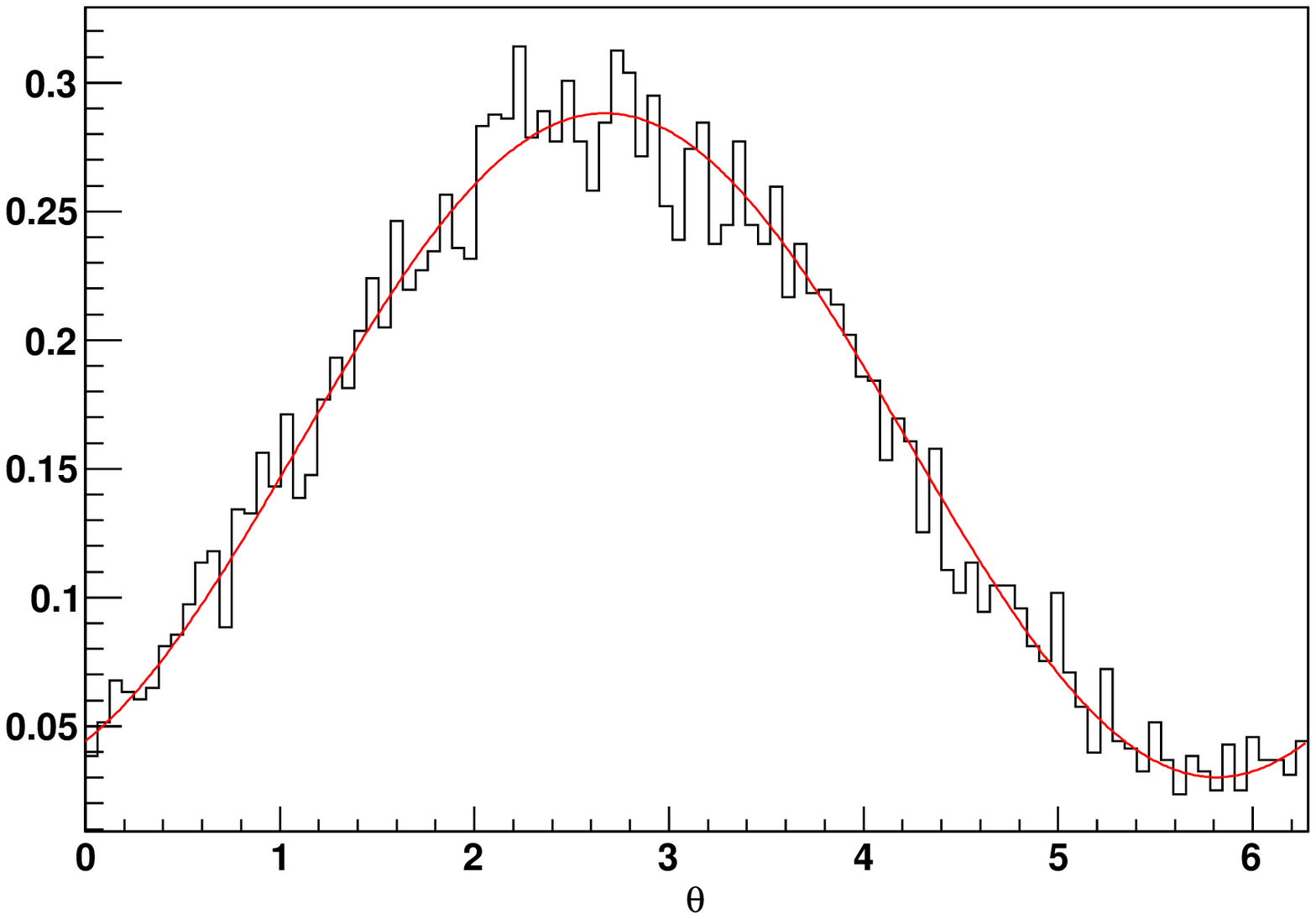}
 \includegraphics[width=0.32\textwidth]{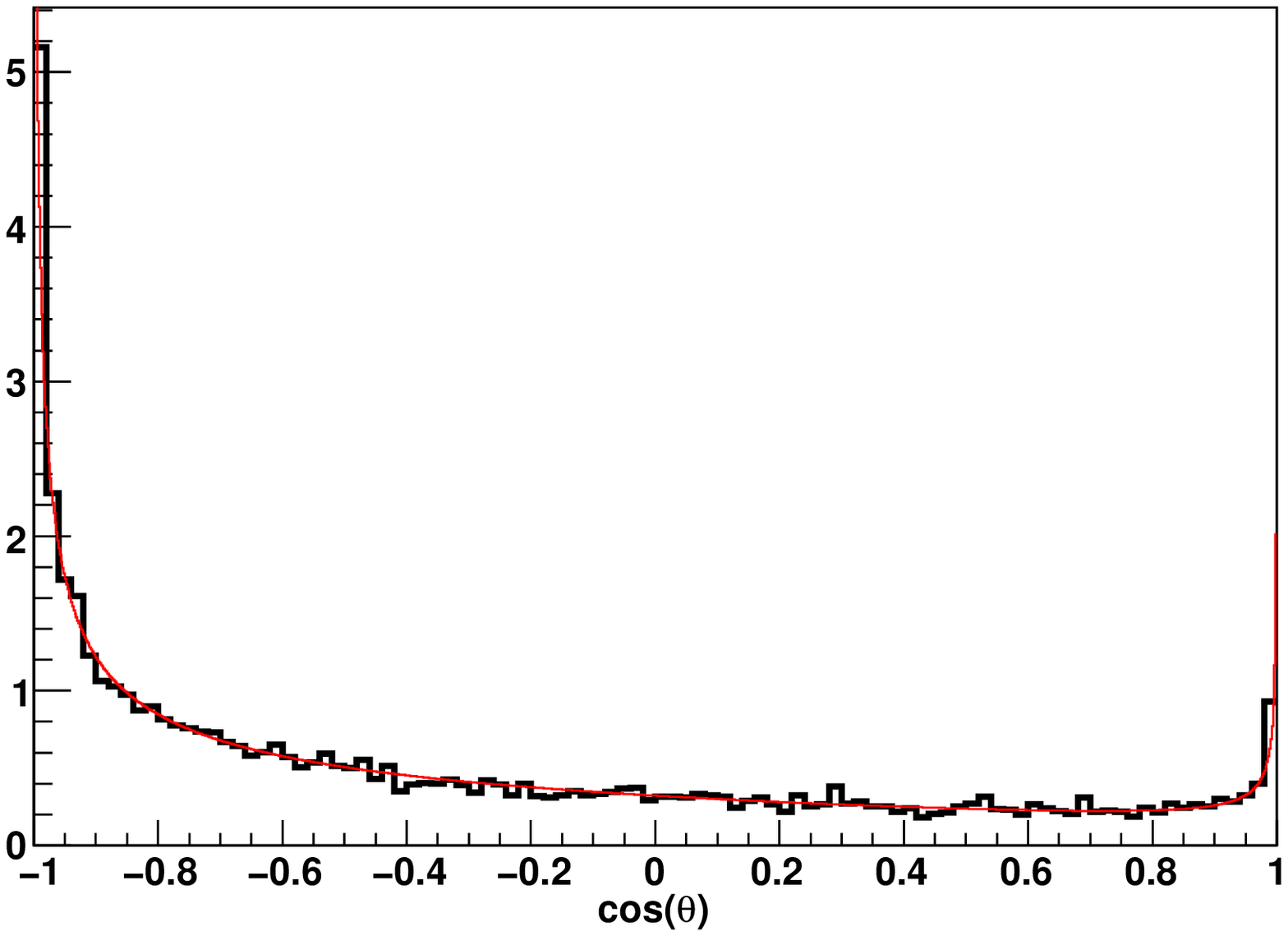}
 \includegraphics[width=0.32\textwidth]{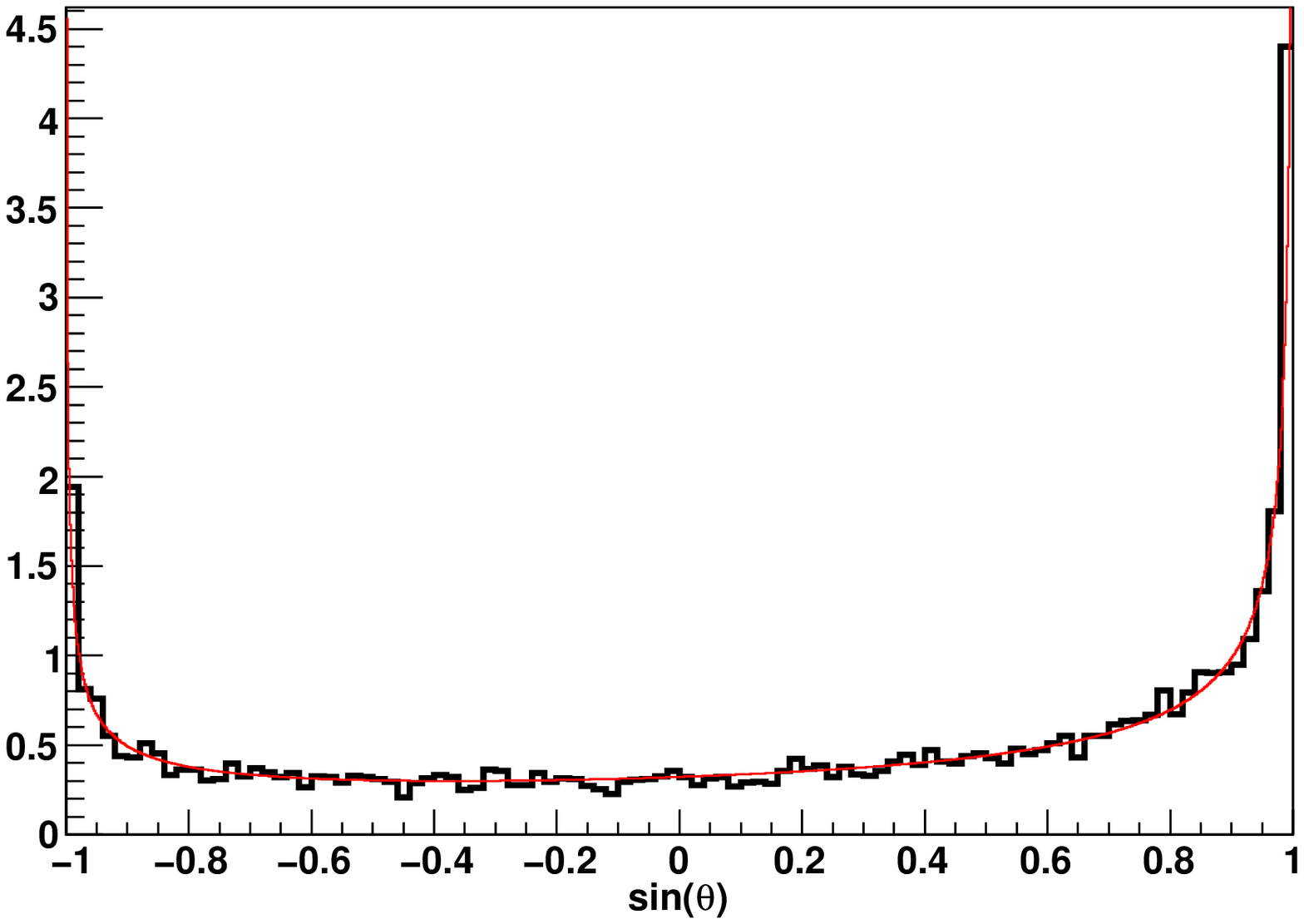}
 \caption{From the simulation of an observation lasting $T_{\rm obs}$ = 3 hours, are showed the distributions of $\theta$ (left), cos$(\theta)$ (middle), and sin$(\theta)$ (right) for $\omega = \omega_0 + 0.35 \cdot 2\pi/T_{\rm obs}$.
\label{SimulationPanels2}}
\end{figure*}

Figure \ref{sc@peak} shows the distributions of $\cos(\theta)$ and $\sin(\theta)$ (left and right panel, respectively) obtained for $\omega=\omega_0$. In this case $P_{\theta}(\theta) = 1 + {\rm sin}(\theta)$, and it allows for a simplification of  the terms $2\pi(i + k)$  in Eqs. (\ref{Pcos}, \ref{Psin}). The pdf $P_{\rm cos}$ and $P_{\rm sin}$, depicted in red on the plots of Figure \ref{sc@peak} are equal to
\begin{eqnarray}
P_{\rm cos}(\theta) &=& \frac{1}{{\rm sin}(\theta)} \Rightarrow \left\langle {\rm cos}(\theta) \right\rangle = 0  , \\
P_{\rm sin}(\theta) &=& \frac{1+{\rm sin}(\theta)}{{\rm cos}(\theta)} \Rightarrow \left\langle {\rm sin}(\theta) \right\rangle = 0.5.
\end{eqnarray}
More generally, for $\omega \neq \omega_0$, the phase distribution is equal to
\begin{equation}
 P_{\theta}(\theta) = 1+{\rm sin}(\frac{\omega_0}{\omega} \theta),
 \label{9}
\end{equation}
and Eqs. (\ref{Pcos}, \ref{Psin}) can not be simplified. Specifically, when the terms $B(\theta; R)$ are not negligible it is particularly hard to analytically reproduce the pdf $P_{\rm cos}$ and $P_{\rm sin}$, because they have different points of discontinuity. In Figure \ref{SimulationPanels}, we choose these cases that most severely test our formulae.  The phase distribution folded in $2\pi$ is plotted in the left panels of the figure. 
Middle and right panels plot the pdf of cos$(\theta)$ and sin$(\theta)$, respectively. Each row in the figure corresponds to a different simulation with observation time $T_{\rm obs}$ chosen such that $N$ from Eq. (\ref{11}) is equal to zero or one, and the phases are calculated for different $\omega \neq \omega_0$, as specified in the caption. The red line in each plot correspond to the analytical solutions. The perfect agreement with the simulations also in reproducing the discontinuities validates our computations in the previous Section.
In the search of pulsations, the condition $N\gg 1$ is always satisfied, even for only few hours of observations. In this most common case, then, the distributions we plotted appear smoother (as shown in Figure \ref{SimulationPanels2}), and the term $B(\theta; R)$ is completely negligible, because its weight is $1/N \sim 0$.

\subsection{Power spectrum features}

Now we are going to demonstrate that the power spectrum of a sinusoidal signal observed for a finite time has a sinc squared shape ($[{\rm sin}(x)/x]^2$) centered on its proper frequency. 
With this aim, the power spectrum calculated by Eq. (\ref{pow2}) needs the evaluation of the average values of sin$(\theta)$ and cos$(\theta)$,
\begin{eqnarray}
\left\langle {\rm sin}(\theta) \right\rangle &=& \int{\left[ {\rm sin}(\theta)P_{\rm sin}(\theta) \right] {\rm cos}(\theta)d\theta }  , \label{20}
\\
\left\langle {\rm cos}(\theta) \right\rangle &=& \int{ - \left[ {\rm cos}(\theta)P_{\rm cos}(\theta) \right] {\rm sin}(\theta)d\theta } .  \label{21}
\end{eqnarray}
The pdf $P_{\rm cos}(\theta)$ and $P_{\rm sin}(\theta)$ are given by Eqs. (\ref{Pcos}) and (\ref{Psin}), respectively. As we already underlined, they are composed by sums of the terms $P_{\theta}(2\pi(i+k) \pm \theta)$. For a sinusoidal signal, which general distribution of the phases is given by Eq. (\ref{9}), these terms are equal to
\begin{equation}
P_{\theta}(2\pi(i+k) \pm \theta) = 1+{\rm sin}\left( 2\pi i \frac{\omega_0}{\omega} + 2\pi k \frac{\omega_0}{\omega} \pm \frac{\omega_0}{\omega} \theta \right) .
\label{22}
\end{equation}
The power will be not null at frequencies close enough to the signal, such that $\omega_0/\omega \sim 1$. 
This approximation can be adopted in Eq.~(\ref{22}), but just for the terms that do not contain the integer $i$, which is the index of the sums in Eqs. (\ref{Pcos}) and (\ref{Psin}). Indeed, for large values of $i$, the term $2\pi i \omega_0/\omega$ can significantly differ by an integer multiple of $2\pi$, and so can not be simplified. Eq. (\ref{22}) is then approximately equal to
\begin{eqnarray}
P_{\theta}(2\pi(i+k) \pm \theta) \sim 1+{\rm sin}\left( 2\pi i \frac{\omega_0}{\omega} + 2\pi k \pm \theta \right) =
\nonumber \\
 1 \pm {\rm sin}\left( 2\pi i \frac{\omega_0}{\omega}\right)  {\rm cos}(\theta)  \pm {\rm cos}\left( 2\pi i \frac{\omega_0}{\omega}\right)  {\rm sin}(\theta).
\label{23}
\end{eqnarray}

To calculate $\left\langle {\rm sin}(\theta) \right\rangle$, only the term containing sin$(\theta)$ in Eq. (\ref{23}) 
will lead to a not null quantity when substituted in Eq. (\ref{Psin}), 
and this one into Eq. (\ref{20}). The average value of sin$(\theta)$ is then equal to
\begin{eqnarray}
\left\langle {\rm sin}(\theta) \right\rangle = \frac{1}{2} \left[ \frac{1}{N} \sum_{i=0}^{N-1} {\rm cos}(\frac{\omega_0}{\omega}2\pi i) \right] ,
\label{24}
\end{eqnarray}
where $1/N$ come from the normalization of $P_{\rm sin}(\theta)$.
For a random value of the ratio $\omega_0/\omega$ the former sum is negligible,
but when the ratio is close to 1 we can write it as
\begin{eqnarray}
\frac{\omega_0}{\omega} = 1 - \frac{\Delta \omega}{\omega},
\label{24b}
\end{eqnarray}
and all the values in the sum will be positive until $2\pi N |\Delta \omega|/\omega \leq \pi/2$, which neglecting $R$ in Eq. (\ref{11}) becomes
\begin{eqnarray}
|\Delta \omega| \leq \frac{1}{4} \omega_T ,
\label{25}
\end{eqnarray}
where $\omega_T = 2\pi/T_{\rm obs}$. 
With good approximation, the term in square brackets in Eq. (\ref{24}) 
is equal to the following integral expression
\begin{eqnarray}
 \frac{1}{N} \sum_{i=0}^{N-1} {\rm cos}(\frac{\omega_0}{\omega}2\pi i) \rightarrow \frac{1}{2\pi N \varepsilon}\int_{0}^{2\pi N \varepsilon} {{\rm cos}(x) dx} \nonumber \\ \hspace{2cm}  = \frac{{\rm sin}(2\pi N \varepsilon)}{2\pi N \varepsilon}, 
 \label{26}
\end{eqnarray}
with $\varepsilon = \Delta \omega/\omega$.

\begin{figure}
\center
\includegraphics[width=0.75\textwidth]{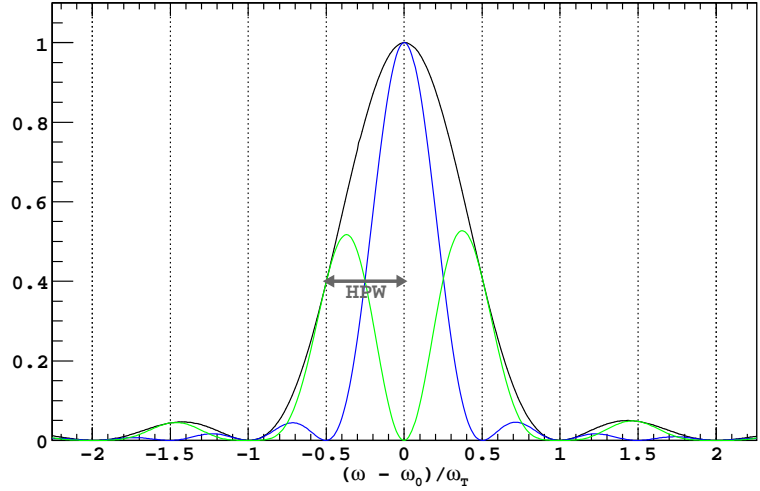}
\caption{Black: power spectrum of a sinusoidal signal. Blue: the contribution of the sine sum $\left( \sum {\rm sin}(\theta_i)\right) ^2$ to the power spectrum. Green: the contribution of the cosine sum $\left( \sum {\rm cos}(\theta_i)\right) ^2$. The half width of the peak (HPW) is also indicated by the double arrow. \label{SingleP}}
\end{figure}

In the same way, the average value of cos$(\theta)$ is evaluated substituting Eq. (\ref{23}) in 
Eq. (\ref{Pcos}), and this one into Eq. (\ref{21}), leading to
\begin{eqnarray}
\left\langle {\rm cos}(\theta) \right\rangle =  - \frac{1}{2} \left[ \frac{1}{N} \sum_{i=0}^{N-1} {\rm sin}(\frac{\omega_0}{\omega}2\pi i) \right]  .
\label{27}
\end{eqnarray}
In this case, all the terms in the sum have the same sign when $2\pi N |\Delta \omega|/\omega \leq \pi$. This condition is less constraining with respect to Eq. (\ref{25}), and can be adopted to define the half peak width ($HPW$) in the power spectrum around $\omega_0$
\begin{eqnarray}
HPW = \frac{1}{2} \omega_T .
\end{eqnarray}
The integral expression for the term in square brackets in Eq. (\ref{27}) is
\begin{eqnarray}
\frac{1}{N} \sum_{i=0}^{N-1} {\rm sin}(\frac{\omega_0}{\omega}2\pi i) \rightarrow \frac{1}{2\pi N \varepsilon}\int_{0}^{2\pi N \varepsilon} {{\rm sin}(x) dx} \nonumber \\ \hspace{2cm} = \frac{1-{\rm cos}(2\pi N \varepsilon)}{2\pi N \varepsilon} .
\label{29}
\end{eqnarray}
From Eq. (\ref{pow2}), the power spectrum is calculated adding the squares of the sine and cosine averages.
Correspondingly, from Eqs. (\ref{26}) and (\ref{29}) we have
\begin{eqnarray}
\left[ \frac{{\rm sin}(2\pi N \varepsilon)}{2\pi N \varepsilon} \right]^2  + \left[ \frac{1-{\rm cos}(2\pi N \varepsilon)}{2\pi N \varepsilon} \right]^2 = \left[ \frac{{\rm sin}(\pi N \varepsilon)}{\pi N \varepsilon} \right]^2
\label{30}
\end{eqnarray}
where $\pi N \varepsilon = \pi \Delta \omega / \omega_T$.
Figure \ref{SingleP} shows the power spectrum (black curve) calculated from Eq. (\ref{pow}) 
for a sinusoidal signal, centered at its proper frequency ($\omega_0$), and in units of $\omega_T$. 
The contribution of the sine sum is shown in blue $\left( \sum {\rm sin}(\theta_i)\right) ^2$ and that of the cosine sum is shown in green $\left( \sum {\rm cos}(\theta_i)\right) ^2$, which are equal to the first and second term on the left hand of Eq. (\ref{30}), respectively. The right hand term of Eq. (\ref{30}) is a squared sinc function centered on the signal frequency, and with width inversely proportional to the observation time. 

In Figure \ref{SingleP} the power spectrum is normalized so that the peak is equal to 1. 
We have considered in this demonstration a  100\% pulsed sinusoidal signal (see Eq. (\ref{8})). 
In contrast, a signal partially pulsed can be represented by the following distribution of the arrival times
\begin{equation}
 P_t(t) = 1 + a \, {\rm sin}(\omega_0 t),
 \label{fracP}
\end{equation}
where $0 \leq a \leq 1$ determines the fraction of the signal that is pulsed.
Then, in Eqs. (\ref{24}) and (\ref{27}) the term multiplying the square brackets is $a/2$,
which substituting in Eq. (\ref{pow2}) results in the power spectrum being proportional to $a^2$.
The peak power in Figure \ref{SingleP} would be equal to $a^2$.
On the other hand, the mean power at frequencies far away from $\omega_0$ remains unchanged. Then, the signal to noise ratio in the power spectrum is proportional to $a^2$. Specifically, it is 
$P(\omega_0)/\left\langle P(\omega\neq\omega_0)\right\rangle = N_0 a^2/4$.

\section{Pulsar frequency derivative}

We are going to apply the method developed in this paper to the practical case of observations so long that the first derivative of the pulsar frequency can not be neglected.
The time series of the emitted photons ($t_e$) by an isolated pulsar can be corrected for the first frequency derivative as:
\begin{equation}
\omega_0 t_c = \omega_0 t_e + \frac{1}{2} \dot{\omega}_0 t_e^2 ,
\label{Fdot-1}
\end{equation}
where $t_c$ is the corrected time series, and $\dot{\omega}_0$ is the frequency derivative. 
If the frequency derivative is un-known, one should try different values of $\dot{\omega}$, which will affect the correction of time series
\begin{equation}
\omega_0 t_w = \omega_0 t_e + \frac{1}{2} \dot{\omega} t_e^2 .
\label{Fdot-2}
\end{equation}
Here $t_w$ stays for generally corrected time series, while $t_c$ is the properly corrected time series. Anyway, once corrected the phase assigned to each photon is
\begin{equation}
\theta = \omega_0 t_w .
\label{Fdot-6}
\end{equation}
In order to apply our method, we need to evaluate the distribution of the phases $P_{\theta}$. 
With this aim we have first to find the relationship between $t_c$ and $t_w$.
From Eqs. \ref{Fdot-1} and \ref{Fdot-2}
\begin{equation}
\omega_0 t_w = \omega_0 t_c + \frac{1}{2} \delta\dot{\omega} t_e^2 ,
\label{Fdot-3}
\end{equation}
where $\delta\dot{\omega} = \dot{\omega}-\dot{\omega}_0$.
Solving Eq. \ref{Fdot-2} for $t_e$ we have
\begin{equation}
t_e = - \frac{\omega_0}{\dot{\omega}} \left[ 1 - \sqrt{1+2\frac{\dot{\omega}}{\omega_0}t_w} \right] ,
\label{Fdot-4}
\end{equation}
where we choose the solution with the negative sign of the square root because this satisfies the condition that $t_e = 0$ when  $t_w=0$. 
Squaring Eq. \ref{Fdot-4}, expanding the root square in the Taylor series until the third term ($\dot{\omega}/\omega_0 t_w \ll 1$ for all the pulsars), 
and substituting in Eq. \ref{Fdot-3} we have
\begin{equation}
t_c \sim t_w -\frac{1}{2} \frac{\delta\dot{\omega}}{\omega_0} t_w^2 . 
\label{Fdot-5}
\end{equation}
and its inverse
\begin{equation}
t_w = \frac{1- \sqrt{1-2\frac{\delta\dot{\omega}}{\omega_0} t_c}}{\frac{\delta\dot{\omega}}{\omega_0}} .
\label{Fdot-7}
\end{equation}

The distribution of the corrected time series $P_{t_w}$ can be calculated applying the formula in Eq. \ref{rotondi}. Since $t_w$ is a monotonic function of $t_c$, the evaluation of $P_{t_w}$ is simplified as
\begin{equation}
P_{t_w}(t_c) = U \frac{P_{t_c}(t_c)}{dt_w/dt_c} , 
\label{Fdot-8}
\end{equation}
where here --and hereafter-- $U$ indicates a normalisation factor.
In the same way, the distribution of the phase assigned to each photon can be caculated considering Eq. \ref{Fdot-6}. 
Since in Eq. \ref{Fdot-6} $\omega_0$ acts like a constant, $P_{\theta}$ has the same form as $P_{t_w}$
\begin{equation}
P_{\theta}(t_c) = U \frac{P_{t_c}(t_c)}{dt_w/dt_c} . 
\label{Fdot-9}
\end{equation}
We assume that the properly corrected times have a sinusoidal distribution
\begin{equation}
P_{t_c}(t_c) = 1 + {\rm sin}(\omega_0 t_c) .
\label{Fdot-10}
\end{equation}
Substituting in Eq. \ref{Fdot-9}
\begin{equation}
P_{\theta}(t_c) = U \frac{1 + {\rm sin}(\omega_0 t_c)}{1/\sqrt{1-2\frac{\delta\dot{\omega}}{\omega_0} t_c}}
\label{Fdot-11}
\end{equation}
The square root at the denominator can be approximated to one, since $2\frac{\delta\dot{\omega}}{\omega_0} t_c \ll 1$ for all the pulsars even for observations as long as some years. Then, substituting $t_c$ with Eq. \ref{Fdot-5} in the argument of the sine we have
\begin{equation}
P_{\theta}(t_w) \sim U \left[ 1 + {\rm sin}(\omega_0 t_w - \frac{1}{2} \delta\dot{\omega} t_w^2) \right] .
\label{Fdot-12}
\end{equation}
Finally, the distribution $P_{\theta}$ as function of $\theta$ is obtained substituting $t_w = \theta / \omega_0$
\begin{equation}
P_{\theta}(\theta) \sim U \left[ 1 + {\rm sin}(\theta - \frac{1}{2} \frac{\delta\dot{\omega}}{\omega_0^2} \theta^2) \right] .
\label{Fdot-13}
\end{equation}

We should substitute Eq. \ref{Fdot-13} in Eqs. \ref{Pcos} and \ref{Psin} to calculate $P_{\rm cos}(\theta)$ and $P_{\rm sin}(\theta)$, which are composed by sums of the terms $P_{\theta}(2\pi (i+k) \pm \theta)$. In this case these terms are equal to
\begin{equation}
P_{\theta}(2\pi (i+k) \pm \theta) = 1 + {\rm sin}(2\pi i + 2\pi k \pm \theta - \frac{1}{2} \frac{\delta\dot{\omega}}{\omega_0^2} (2\pi i + 2\pi k \pm \theta)^2 ) .
\label{Fdot-14}
\end{equation}
Since here $0 \leqslant \theta < 2\pi$ while $2\pi i$ can be as large as $\omega_0 T_{obs}$ (see Eq. 3.8), then in the squared term $\pm \theta$ can be neglected. Thus
\begin{eqnarray}
P_{\theta}(2\pi (i+k) \pm \theta) \sim 1 + {\rm sin}(2\pi k \pm \theta - 2\pi^2 \frac{\delta\dot{\omega}}{\omega_0^2} (i + k)^2 ) = \nonumber \\
1 \pm {\rm sin}(2\pi^2 \frac{\delta\dot{\omega}}{\omega_0^2} (i + k)^2){\rm cos}(\theta) \pm {\rm cos}(2\pi^2 \frac{\delta\dot{\omega}}{\omega_0^2} (i + k)^2){\rm sin}(\theta) .
\label{Fdot-15}
\end{eqnarray}
To calculate $\left\langle {\rm sin} \theta \right\rangle$ , only the term containing sin$(\theta)$ in Eq. (\ref{Fdot-15}) will lead to a not null quantity
when substituted in Eq. (\ref{Psin}), and this one into Eq. (\ref{20}). The average value of sin$(\theta)$ is then equal
to
\begin{equation}
\left\langle {\rm sin} \theta \right\rangle = \frac{1}{2}\left[ \frac{1}{N} \sum_{i=0}^{N-1} {\rm cos}(2\pi^2 \frac{\delta\dot{\omega}}{\omega_0^2} (i + k)^2)\right] 
\label{Fdot-16}
\end{equation}
$k = 0, 1/2, 1$ can be neglected.
With good approximation, the term in square brackets in Eq. (\ref{Fdot-16}) is equal to
the following integral expression
\begin{equation}
\frac{1}{N} \sum_{i=0}^{N-1} {\rm cos}\left( 2\pi^2 \frac{\delta\dot{\omega}}{\omega_0^2} i^2\right)  \rightarrow 
\frac{1}{\sqrt{\pi y}} \int_{0}^{\sqrt{\pi y}} {{\rm cos}(x^2) dx} ,
\label{Fdot-17}
\end{equation}
where
\begin{equation}
y = 2 \pi N^2 \frac{\delta\dot{\omega}}{\omega_0^2} = 2 \pi \frac{\delta\dot{\omega}}{\omega_T^2} .
\label{Fdot-18}
\end{equation}
Then
\begin{equation}
\left\langle {\rm sin} \theta \right\rangle = \frac{1}{2} C(\sqrt{\pi y})
\label{Fdot-19}
\end{equation}
where $C(x) = \int_{0}^{x} {{\rm cos}(t^2) dt}$ is the cosine Fresnel integral.

In the same way, the average value of cos$(\theta)$ is evaluated substituting Eq. (\ref{Fdot-15}) in 
Eq. (\ref{Pcos}), and this one into Eq. (\ref{21}), leading to
\begin{eqnarray}
\left\langle {\rm cos}(\theta) \right\rangle =  - \frac{1}{2} \left[ \frac{1}{N} \sum_{i=0}^{N-1} {\rm sin}(2\pi^2 \frac{\delta\dot{\omega}}{\omega_0^2} i^2) \right]  .
\label{Fdot-20}
\end{eqnarray}
All the terms in the sum are positive until $ 2\pi^2 \frac{\delta\dot{\omega}}{\omega_0^2} N^2 \leq \pi$. This condition can be adopted to define the width of the peak in the power spectrum at variance of $\delta\dot{\omega}$
\begin{eqnarray}
HPW_{\delta\dot{\omega}} = \frac{\omega_T^2}{2\pi}  .
\label{Fdot-21}
\end{eqnarray}
The integral expression for the term in square brackets in Eq. (\ref{Fdot-20}) is
\begin{equation}
\frac{1}{N} \sum_{i=0}^{N-1} {\rm sin}\left( 2\pi^2 \frac{\delta\dot{\omega}}{\omega_0^2} i^2\right)  \rightarrow 
\frac{1}{\sqrt{\pi y}} \int_{0}^{\sqrt{\pi y}} {{\rm sin}(x^2) dx} .
\label{Fdot-22}
\end{equation}
Then
\begin{equation}
\left\langle {\rm cos} \theta \right\rangle = \frac{1}{2} S(\sqrt{\pi y})
\label{Fdot-23}
\end{equation}
where $S(x) = \int_{0}^{x} {{\rm sin}(t^2) dt}$ is the sine Fresnel integral.

\begin{figure}
\center
\includegraphics[width=0.75\textwidth]{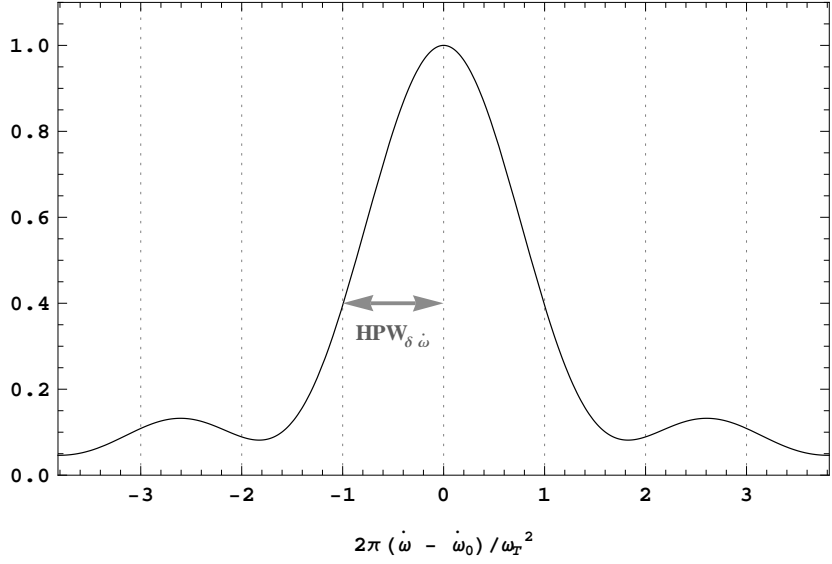}
\caption{Power spectrum of a sinusoidal signal at variance of $\delta\dot{\omega}$ in units of $2\pi \delta\dot{\omega} / \omega_T^2$. 
\label{FdotShape}}
\end{figure}

From Eq. (\ref{pow2}), the power spectrum is calculated adding the squares of the sine and cosine averages.
Correspondingly, from Eqs. (\ref{Fdot-19}) and (\ref{Fdot-23}) we have
\begin{eqnarray}
P(\omega) = U \left[ \left\langle {\rm sin} \theta \right\rangle^2 + \left\langle {\rm cos} \theta \right\rangle^2 \right] =
U\frac{S(\sqrt{\pi y})^2 + C(\sqrt{\pi y})^2} {\pi y}.
\label{Fdot-24}
\end{eqnarray}
where $U$ is a normalization factor, which in Figure \ref{FdotShape} is choosen so that the power peak is equal to 1.
Figure \ref{FdotShape} shows the shape of the power spectrum at variance of $\delta\dot{\omega}$ as function of the variable $y = 2\pi \delta\dot{\omega} / \omega_T^2 $. In these units the width of the peak is equal to $y=1$, and the first minimum is at $y \sim 1.8$. Figure \ref{FdotShape} and Eq. (\ref{Fdot-21}) show that in a pulsation search the first frequency derivative can not be neglected when $\dot{\omega}_0$ is of the order of magnitude of $1/T_{obs}^2$, or greater.

\section{Blind search}

A more general case happens when both the pulsar frequency and its first derivative are unknown.
Of course, this happen every time one search for new pulsars, 
but there are at least two situations where the first frequency derivative can not be neglected in the search. On the one hand, 
the search for radio quiet $\gamma$-ray pulsars with $\gamma$-ray data needs integration times of few weeks or more, so that 
$\dot{\omega}_0$ is not negligible. 
On the other hand, in radio searches for pulsars with fast spin down $\dot{\omega}_0$, it is important even for observations of few hours.

The general form of Eqs. (\ref{Fdot-1}), and (\ref{Fdot-2}) is:
\begin{eqnarray}
t_c = t_e + \frac{1}{2} \frac{\dot{\omega}_0}{\omega_0} t_e^2 
\label{B1} \\
t_w = t_e + \frac{1}{2} \frac{\dot{\omega}}{\omega} t_e^2.
\label{B2}
\end{eqnarray}
Following the same steps and approximations from Eq. (\ref{Fdot-3}) to Eq. (\ref{Fdot-6}), and from Eq. (\ref{Fdot-8}) to Eq. (\ref{Fdot-13}) we get for the general case
\begin{eqnarray}
&& t_c = t_w - \frac{\delta R}{2} t_w^2 
\label{B3} \\ &&
P_{\theta}(\theta) = U \left[ 1 + {\rm sin}\left( \frac{\omega_0}{\omega} \theta - \frac{\delta R}{2} \frac{\omega_0}{\omega^2}  \theta^2\right)  \right] .
\label{B4}
\end{eqnarray}
where 
\begin{eqnarray}
\delta R = \frac{\dot{\omega}}{\omega} - \frac{\dot{\omega}_0}{\omega_0} \sim 
\frac{1}{\omega_0^2}(\omega_0 \delta \dot{\omega} - \dot{\omega}_0 \Delta \omega) .
\label{B5}
\end{eqnarray}
To calculate $P_{\rm cos}(\theta)$, and $P_{\rm sin}(\theta)$ we substitute in Eq. (\ref{B4}) $\theta$ with $2 \pi (i+k) \pm \theta$, and we apply the same approximations as in Eq. (\ref{Fdot-15}) (neglecting $k$ and $\theta$ when possible):
\begin{eqnarray}
P_{\theta}(2\pi (i+k) \pm \theta) \sim 1 + {\rm sin}(\frac{\omega_0}{\omega} (2\pi i \pm \theta) - \delta R \frac{\omega_0}{\omega^2} 2\pi^2 i^2 ).
\label{B6}
\end{eqnarray}
The first term within the sine is equal to
\begin{eqnarray}
\frac{\omega_0}{\omega} (2\pi i \pm \theta) = \left( 1-\frac{\Delta \omega}{\omega}\right) (2\pi i \pm \theta) \sim 
2\pi i \pm \theta - \frac{\Delta \omega}{\omega} 2\pi i.
\label{B7}
\end{eqnarray}
Substituting in Eq. (\ref{B5}) we have
\begin{eqnarray}
&& P_{\theta}(2\pi (i+k) \pm \theta) \sim
\nonumber \\ &&
1 \pm {\rm sin}(\frac{\Delta \omega}{\omega} 2\pi i + \delta R \frac{\omega_0}{\omega^2} 2\pi^2 i^2){\rm cos}(\theta) \pm {\rm cos}(\frac{\Delta \omega}{\omega} 2\pi i + \delta R \frac{\omega_0}{\omega^2} 2\pi^2 i^2){\rm sin}(\theta) .
\label{B8}
\end{eqnarray}
Only the term multiplying sin$(\theta)$ in Eq. (\ref{B8}) gives a not null contribution to $\left\langle {\rm sin}(\theta) \right\rangle$
\begin{eqnarray}
\left\langle {\rm sin}(\theta) \right\rangle = \frac{1}{2}\left[ \frac{1}{N}\sum_{i=0}^{N-1} {\rm cos}\left( \frac{2\pi i}{\omega}\left[ \Delta \omega + \frac{\delta R}{2} \omega_0 \frac{2\pi i}{\omega}\right] \right) \right] . 
\label{B9}
\end{eqnarray}
Setting $z = 2\pi i/\omega$, the term within the square brackets can be approximated with
\begin{eqnarray}
&&  \frac{1}{N}\sum_{i=0}^{N-1} {\rm cos}\left( \frac{2\pi i}{\omega}\left[ \Delta \omega + \frac{\delta R}{2} \omega_0 \frac{2\pi i}{\omega}\right] \right) \rightarrow 
\frac{\omega}{2 \pi N} \int_{0}^{2\pi N/\omega} {{\rm cos}\left(\Delta \omega z + \frac{\delta R}{2} \omega_0 z^2 \right) dz} =
\nonumber \\ &&
\frac{\omega}{2N \sqrt{\pi \delta R \omega_0} } \left\lbrace  {\rm cos} \left( \frac{\Delta \omega^2}{2 \delta R \omega_0} \right) 
\left[ -C\left( \frac{\Delta \omega}{\sqrt{\pi \delta R \omega_0}} \right) + C\left( \frac{\Delta \omega + 2N\pi \delta R 
\omega_0/\omega}{\sqrt{\pi \delta R \omega_0}}\right) \right] +\right.
\nonumber \\ && \left.
{\rm sin} \left( \frac{\Delta \omega^2}{2 \delta R \omega_0} \right) 
\left[ -S\left( \frac{\Delta \omega}{\sqrt{\pi \delta R \omega_0}} \right) + S\left( \frac{\Delta \omega + 2N\pi \delta R \omega_0/\omega}{\sqrt{\pi \delta R \omega_0}}\right)   \right]  \right\rbrace 
\label{B10}
\end{eqnarray}
where $S(x)$ and $C(x)$ are the sine and cosine Fresnel integrals, respectively.
Similarly, the average value of cos$(\theta)$ is 
\begin{eqnarray}
\left\langle {\rm cos}(\theta) \right\rangle = \frac{1}{2}\left[ \frac{1}{N}\sum_{i=0}^{N-1} {\rm sin}\left( \frac{2\pi i}{\omega}\left[ \Delta \omega + \frac{\delta R}{2} \omega_0 \frac{2\pi i}{\omega}\right] \right) \right] , 
\label{B11}
\end{eqnarray}
and the term within the square brackets can be approximated with
\begin{eqnarray}
&&  \frac{1}{N}\sum_{i=0}^{N-1} {\rm sin}\left( \frac{2\pi i}{\omega}\left[ \Delta \omega + \frac{\delta R}{2} \omega_0 \frac{2\pi i}{\omega}\right] \right) \rightarrow 
\frac{\omega}{2 \pi N} \int_{0}^{2\pi N/\omega} {{\rm sin}\left(\Delta \omega z + \frac{\delta R}{2} \omega_0 z^2 \right) dz} =
\nonumber \\ &&
\frac{\omega}{2N \sqrt{\pi \delta R \omega_0} } \left\lbrace  {\rm cos} \left( \frac{\Delta \omega^2}{2 \delta R \omega_0} \right) 
\left[ -S\left( \frac{\Delta \omega}{\sqrt{\pi \delta R \omega_0}} \right) + S\left( \frac{\Delta \omega + 2N\pi \delta R 
\omega_0/\omega}{\sqrt{\pi \delta R \omega_0}}\right) \right] -\right.
\nonumber \\ && \left.
{\rm sin} \left( \frac{\Delta \omega^2}{2 \delta R \omega_0} \right) 
\left[ -C\left( \frac{\Delta \omega}{\sqrt{\pi \delta R \omega_0}} \right) + C\left( \frac{\Delta \omega + 2N\pi \delta R \omega_0/\omega}{\sqrt{\pi \delta R \omega_0}}\right)   \right]  \right\rbrace .
\label{B12}
\end{eqnarray}
Before writing the formula of the expectation value of the power spectrum is useful to make the following simplifications.
\begin{eqnarray}
&& \left[ \frac{2N \sqrt{\pi \delta R \omega_0}}{\omega} \right]^2 = 
\frac{4\pi}{\omega_T^2 \omega_0}(\omega_0 \delta \dot{\omega} - \dot{\omega_0} \Delta \omega) =
\nonumber \\
&& \frac{4\pi \delta \dot{\omega}}{\omega_T^2} - \frac{4\pi}{\omega_T} \frac{\dot{\omega_0}}{\omega_0} \frac{\Delta \omega}{\omega_T} = 2y - Kx ,
\label{B13}
\end{eqnarray}
where from Eq. (\ref{11}) $N=\omega/\omega_T$, $\delta R$ is given by Eq. (\ref{B5}), and we set
\begin{eqnarray}
&& y = \frac{2\pi \delta \dot{\omega}}{\omega_T^2}
\label{B14} \\ 
&& x = \frac{\Delta \omega}{\omega_T}
\label{B15} \\ 
&& K = \frac{4\pi}{\omega_T} \frac{\dot{\omega_0}}{\omega_0} .
\label{B16}
\end{eqnarray}
With this notation the arguments of the Fresnel integrals in Eqs. (\ref{B10}), and (\ref{B12}) are
\begin{eqnarray}
&& \frac{\Delta \omega}{\sqrt{\pi \delta R \omega_0}} = \frac{2N\Delta \omega /\omega}{2N \sqrt{\pi \delta R \omega_0}/\omega} 
= \frac{2x}{\sqrt{2y-Kx}}
\label{B17} \\ 
&& \frac{\Delta \omega + 2N\pi \delta R \omega_0/\omega}{\sqrt{\pi \delta R \omega_0}} =
\frac{\Delta \omega}{\sqrt{\pi \delta R \omega_0}} + \frac{2N \sqrt{\pi \delta R \omega_0}}{\omega} =
\frac{2x}{\sqrt{2y-Kx}} + \sqrt{2y-Kx} .
\label{B18} 
\end{eqnarray}
Finally, the power spectrum given by the sum of the squares of Eqs.(\ref{B10}), and (\ref{B12}) is
\begin{eqnarray}
P(x, y) = \frac{1}{2y-Kx} && \left\lbrace \left[ C\left( \frac{2x}{\sqrt{2y-Kx}} \right) - C\left( \frac{2x}{\sqrt{2y-Kx}} + \sqrt{2y-Kx}\right) \right]^2 + \right.
\nonumber \\ && \left.
 \left[ S\left( \frac{2x}{\sqrt{2y-Kx}} \right) - S\left( \frac{2x}{\sqrt{2y-Kx}} + \sqrt{2y-Kx}\right) \right]^2 \right\rbrace .
\label{B19} 
\end{eqnarray}

\begin{figure}
\center
\includegraphics[width=0.75\textwidth]{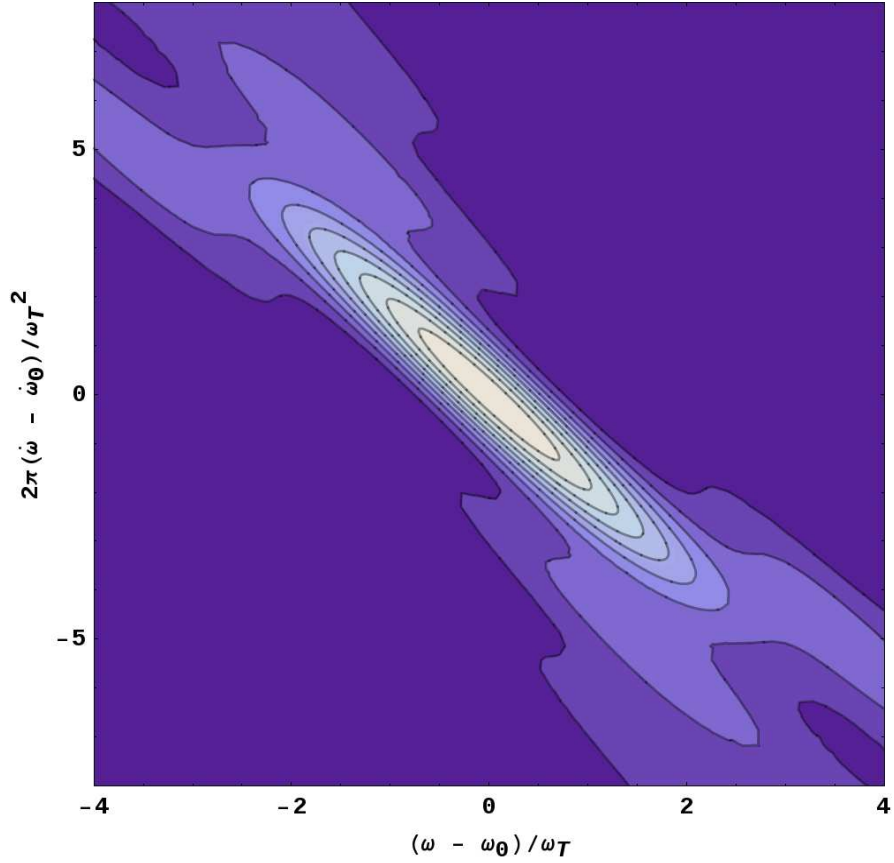}
\caption{Contour plot of the power spectrum of a sinusoidal signal at variance of $\Delta \omega$ and $\delta\dot{\omega}$ in units of $\Delta \omega/\omega_T$, $2\pi \delta\dot{\omega} / \omega_T^2$, respectively. 
The maximum power is equal to 1 at the origin of the axes. The contours range from 0.1 to 0.9 in steps of 0.1.
In this plot the parameter $K$ of Eq. (\ref{B18}) is set $K=0.1$. 
\label{FBlind}}
\end{figure}

Figure \ref{FBlind} shows the power spectrum at variance of both $\Delta \omega$ and $\delta \dot{\omega}$.
The shape of the power spectrum follow an oblique structure, which is the typical one observed in plots produced 
for example by the program {\it PRESTO} when a blind search is performed. The diagonal axis of the structure has a weak dependence by the parameter $K$ of Eq. (\ref{B18}) when it is lower than 1. For $K\rightarrow 0$ the diagonal axis has the equation $y=-2x$, that means 
\begin{eqnarray}
\delta \dot{\omega} = - \Delta \omega \frac{\omega_T}{\pi} .
\label{B20} 
\end{eqnarray}

\section{Conclusions}

In this paper we describe and  validate a method to calculate the expectation value of the power spectrum.
Adopting the definition given by \cite{Scargle1982} (see Eq. \ref{pow}) we calculate the expectation value making use of the statistical properties of the arrival time series, and consequently of the phases attributed to each event. 
Our results are summarized by Eqs. (\ref{pow2}), (\ref{Pcos}), and (\ref{Psin}).

We validate the method focusing on the simple case of a sinusoidal signal assumed to come from an isolated pulsar. But since the solutions in Eq. (\ref{Pcos}) and (\ref{Psin}) are free from any assumption on the event phase distribution, the method can be generalized to any situation.


As noticed at the end of Section 2, a key ingredient of our method is the sum of the terms $\sum_{i=0}^{N-1} P_{\theta}(2\pi(i+k) \pm \theta)$, which corresponds to the distribution of the event phases folded by $2\pi$. At the proper frequency $\omega_0$, the folded distribution is equivalent to the pulse profile, but this is not true anymore at a different frequency $\omega \neq \omega_0$, as shown for example by Eq. (\ref{9}) in the case of a sinusoidal signal. There are several factors that can modify the folded distribution of the phases. 
In this paper we applied our method to the case in which the folded distribution is perturbed by the first derivative of the pulsar frequency. Also, we considered the power spectrum expected in a blind search, in which both the frequency and its first derivative are uncertain. The analytical descriptions of the power spectra in these cases are given by Eq. (\ref{Fdot-24}) and Eq. (\ref{B19}), respectively. These are novel results in the field of timing.
In a separate paper we make direct use of the method developed here to evaluate the effects of the uncertainties of orbital parameters in the timing of pulsars in binary systems \cite{Cali12}.


\section*{Acknowledgments}

This work was supported by the  grants AYA2012-39303, SGR2009-811, and iLINK2011-0303. DFT was additionally supported by a Friedrich Wilhelm Bessel Award of the Alexander von Humboldt Foundation.

\end{document}